\documentclass[letterpaper]{article}
\usepackage[preprint]{aaai2027}
\usepackage[hyphens]{url}
\usepackage{graphicx}
\usepackage{natbib}
\usepackage{caption}
\usepackage{booktabs}
\usepackage{amsmath}
\title{Confidently Wrong, Silently So: Auditing Undetectable Failures of a Deployed On-Device Language Model\thanks{Preprint. Under review.}}
\author{
    Shashwat Pandey\equalcontrib\textsuperscript{\rm 1},
    Satwik Pandey\equalcontrib\textsuperscript{\rm 2},
    Suresh Raghu\textsuperscript{\rm 2}
}
\affiliations{
    \textsuperscript{\rm 1}University of California, Santa Cruz\\
    \textsuperscript{\rm 2}Independent Researcher\\
    spandey7@ucsc.edu, psatwik2711@gmail.com, sureshraghu0706@gmail.com
}
\begin{document}
\maketitle

\begin{abstract}
Aligning deployed language models requires knowing when their outputs can be trusted, yet on-device models now ship to hundreds of millions of devices with no server-side moderation, and the configuration developers can actually deploy is rarely audited independently. We present a reproducible reliability audit of the developer-accessible on-device foundation model, framed as an oversight question: can a user or a resource-constrained developer tell when the model is wrong? Red-teaming it on calibration, confident confabulation on false-premise questions, and over-refusal of benign prompts, we find a \emph{task-asymmetric miscalibration}: its guardrails fail in opposite directions across tasks (confabulating on 69\% of false premises while refusing 18\% of entirely benign inputs), atop a self-reported confidence that is saturated and non-discriminative (AUROC 0.47; ECE 70, worst among comparable small models). Crucially, confident-correct and confident-wrong outputs are \emph{surface-indistinguishable}: a classifier over 15 user-visible features separates them at AUROC only 0.55 (equivalence-confirmed), leaving no signal for oversight at inference time. No cheap single-generation signal flags these failures ($\le$0.68 AUROC), whereas a black-box consistency wrapper requiring no model access recovers reliability (confident confabulation 75\%$\to$3\%; selective accuracy 43\%$\to$83\%) at a tunable cost. We contribute a model-agnostic audit protocol, a surface-indistinguishability test, and released code and frozen evaluation items as reusable infrastructure for auditing deployed models.
\end{abstract}

\section{Introduction}
\label{sec:intro}
On-device language models now run on hundreds of millions of consumer devices, exposed to third-party developers through vendor frameworks and embedded directly in user-facing applications. Unlike server-hosted models, they typically operate without server-side moderation, a fallback model, or a human in the loop: the model's output is what the user sees. This makes their reliability, not just their capability, a safety-relevant property, since aligning a deployed model in practice presupposes knowing when its outputs can be trusted. Yet independent scrutiny has lagged deployment, and the little public evidence that exists usually concerns the configuration a vendor \emph{benchmarks} rather than the one a developer can actually \emph{ship}.

We conduct an independent, black-box audit of the developer-accessible on-device foundation model (the general base model exposed through the public framework, not the private, task-specialized adapter the vendor reports on), using only the public API on commodity hardware. We red-team it on the reliability axes that matter for unmoderated deployment: calibration, confident confabulation on questions built on false premises, and over-refusal of benign inputs.

Three findings emerge. First, the model exhibits a \emph{task-asymmetric miscalibration}: it confidently confabulates on $69\%$ of false-premise questions (abstaining only $31\%$), while refusing $18\%$ of entirely benign summarization inputs, with guardrails miscalibrated in opposite directions and unchanged by the vendor's own permissive mode. Underlying both is a self-reported confidence that is saturated ($\sim$99\%) and non-discriminative (AUROC $0.47$; expected calibration error $70$, worst among comparable small models). Second, and most consequentially, these confident errors are \emph{surface-indistinguishable} from confident-correct outputs: across fifteen user-visible features, a classifier separates them at AUROC only $0.55$ (statistically equivalent to chance), and the model's errors are \emph{less} detectable than any peer's, so a user has no cue with which to protect themselves. Third, this undetectability is not cheaply repaired: no single-generation signal (verbalized confidence, hedging, or response length) flags the errors ($\le 0.68$ AUROC), and the cheap trace signals that work on server-side reasoning models do not transfer to this setting. What does work is a black-box, decode-time consistency wrapper requiring no model access, which cuts confident confabulation from $75\%$ to $3\%$ and raises selective accuracy from $43\%$ to $83\%$ at a tunable cost. Our transportable contribution is a model-agnostic audit protocol and two framings, surface indistinguishability and the cheap-vs-expensive cost-ladder, released together with code and frozen items as reusable infrastructure; these apply to any deployed model, while the specific rates below instantiate them on one widely deployed system, and we treat cross-model and cross-version generalization as a conjecture (Section~\ref{sec:limitations}).

We make four contributions:
\begin{itemize}
\item An independent reliability red-team of the \emph{developer-accessible} on-device model, revealing a task-asymmetric miscalibration and high-confidence confabulation that survive the vendor's permissive guardrail mode (Section~\ref{sec:findings}).
\item A surface-indistinguishability analysis showing, with negligible effect sizes and a TOST equivalence test, that confident-correct and confident-wrong outputs carry no user-visible signal (Section~\ref{sec:surface}).
\item A cost-ladder establishing that no cheap single-generation signal flags these failures, while $O(N)$ consistency does, with the difference significant on two task types (Section~\ref{sec:costladder}).
\item A black-box mitigation requiring no model access, together with released code and frozen evaluation items, that recovers reliability at a tunable cost (Section~\ref{sec:mitigation}).
\end{itemize}

Beyond the specific model, our results argue that reliability audits, and the claims that inform deployment and policy, should target the configuration developers can ship, and that oversight of on-device models like the one we study should not depend on a single forward pass.

\section{Related Work}
\label{sec:relatedwork}
\paragraph{On-device models and their evaluation.}
Compact language models are now shipped on consumer hardware and exposed to developers through vendor frameworks~\citep{applefm2025,gemma3,llama32}. Public evaluation has centered on capability benchmarks and, for closed models, on the vendor-reported configuration. Independent reliability studies of small models exist~\citep{slmaudit}, but they target open checkpoints in the standard research stack rather than the closed, developer-accessible configuration a vendor actually deploys. \emph{Gap:} the deployable on-device configuration, distinct from the privately benchmarked one, has not been independently audited on reliability axes.

\paragraph{Calibration and single-pass uncertainty.}
A large literature seeks cheap signals for when a model is wrong: verbalized confidence~\citep{tian2023,yoon2025}, trace length~\citep{devic2025}, and self-evaluation~\citep{kadavath2022}, alongside more expensive sampling-based estimators such as semantic entropy~\citep{farquhar2024} and self-consistency~\citep{wang2023selfcons}. These signals are developed and validated almost entirely on server-hosted models that emit rich reasoning traces. \emph{Gap:} whether they transfer to deployed on-device models, which are smaller, guided-decoded, and latency-constrained, is untested; we show the cheap single-generation signals do not.

\paragraph{Confident errors, red-teaming, and over-refusal.}
Confident hallucination~\citep{huang2023hallu}, adversarial red-teaming~\citep{perez2022redteam}, and over-refusal of benign prompts~\citep{rottger2024xstest} are each active areas, and consistency-based methods offer black-box error flagging~\citep{manakul2023selfcheck}. \emph{Gap:} none ties confident errors to \emph{user-undetectability}, whether anyone consuming the output can distinguish a confident error from a confident-correct answer, nor evaluates black-box recovery on a deployed on-device model. Our audit fills these gaps: we measure detectability directly (Section~\ref{sec:surface}), test whether cheap signals transfer (Section~\ref{sec:costladder}), and provide an access-free mitigation (Section~\ref{sec:mitigation}).

\section{Audit Setup and Threat Model}
\label{sec:setup}

\paragraph{Model under test and access boundary.}
We audit the on-device foundation model exposed to third-party developers through the FoundationModels framework, specifically \texttt{SystemLanguageModel.default}, a general-purpose base model of roughly 3B parameters, queried on a commodity laptop. We deliberately study \emph{this} configuration because it is the one developers can actually ship: it is distinct from the private, task-specialized adapter (e.g., for summarization) that the vendor benchmarks in its own technical report but does not expose through the public API. Our claims are therefore scoped to the developer-accessible model as exposed by the public FoundationModels API at the time of study; we do not dispute the vendor's reported results for its private configurations, and we return to this access asymmetry in Section~\ref{sec:discussion}. Unless otherwise noted, generation uses the framework's guided (structured) decoding; the single-generation uncertainty probe of Section~\ref{sec:costladder} instead uses the non-guided free-form condition, since guided decoding suppresses the reasoning trace that hedging signals require, an effect we quantify on MGSM~\citep{shi2022languagemodelsmultilingualchainofthought} in the supplementary material (\S\,Guided Decoding Suppresses Reasoning).

\paragraph{Peers and judges.}
For every metric we compare against three openly available small models of comparable scale, served through a common API: Gemma-3-4B-it, Llama-3.2-3B-instruct, and Ministral-3B. Peers are decoded greedily (temperature $0$) for the main comparison, the lowest-variance setting available through the hosted API and the closest practical match to the on-device model's near-deterministic guided decoding, and we additionally re-evaluate all peers under sampled decoding (temperature $0.7$) to confirm the cross-model rankings are not a decoding artifact (supplementary material, \S\,Peer Decoding Robustness); serving-pipeline differences between on-device and hosted API remain the only uncontrolled factor. Faithfulness judgments (Section~\ref{sec:refusal}) use GPT-4o-mini as the primary judge, with a second judge (Llama-3.3-70B) for an agreement audit ($88\%$ agreement); no human faithfulness audit was performed (see Limitations).

\paragraph{Tasks and item sets.}
We use four frozen item sets. \emph{Factual QA} (TriviaQA~\citep{triviaqa} and Global-MMLU~\citep{globalmmlu}) drives the calibration and confident-hallucination analyses. A \emph{false-premise} set (110 unanswerable items carrying a false presupposition, plus 150 answerable controls) drives the confabulation and abstention analyses. A \emph{summarization} set (310 items spanning email, news, technical Q\&A, and ML news) drives the groundedness and over-refusal analyses. For the single-generation uncertainty probe we additionally use free-form TriviaQA and GSM8K~\citep{gsm8k} with frozen, disjoint calibration/evaluation splits. All items are held fixed and identical across models; item sets are released.

\paragraph{Metrics and protocol.}
All models are scored by one deterministic grader on identical items. We report expected calibration error (ECE), discrimination via AUROC (how well a signal separates correct from incorrect answers), abstention and over-refusal rates, confident-hallucination and confident-confabulation rates, judge-rated groundedness, and selective accuracy/coverage under abstention. Throughout, ``confident'' denotes a verbalized confidence of at least $80/100$. Cross-model differences use paired tests on matched items (bootstrap confidence intervals for rate and AUROC gaps, McNemar's test for paired binary outcomes, DeLong's test for correlated AUROCs, and two one-sided tests, TOST, for equivalence claims in Section~\ref{sec:surface}). Confidence intervals are $95\%$ unless stated otherwise.

\paragraph{Grading abstention and confabulation.}
Grading is fully deterministic. Each false-premise item carries a gold \emph{unanswerable} label; the grader scores a response as an \emph{abstention} if it declines or explicitly challenges the false premise (matched against a released set of refusal/premise-flagging patterns), and as a \emph{confident confabulation} if it instead supplies a substantive answer at verbalized confidence $\ge 80$. Factual correctness uses normalized exact match against gold answers. All grading rules, patterns, and thresholds are released for inspection.

\paragraph{Threat model.}
The party at risk is the end user of an on-device application that surfaces the model's outputs directly, typically without server-side moderation or a fallback model. We treat a \emph{confident error} (a high-confidence output that is factually wrong or that answers a question it should decline) as the harmful event, and we ask two questions: whether such errors are distinguishable from correct outputs by any signal available to the user or a resource-constrained developer (Sections~\ref{sec:surface}--\ref{sec:costladder}), and whether they can be mitigated without privileged model access (Section~\ref{sec:mitigation}).

\section{Findings}
\label{sec:findings}
We organize our findings around a single pattern: the model's reliability guardrails are \emph{miscalibrated in opposite directions depending on the task}. On inputs that warrant caution (questions built on false premises), it fails to abstain and answers confidently (Section~\ref{sec:fp}); on entirely benign inputs, it refuses (Section~\ref{sec:refusal}). Underlying both is a confidence signal that is saturated and non-discriminative (Section~\ref{sec:calib}). All comparisons use one deterministic grader and identical items across models; ``confident'' denotes verbalized confidence $\ge 80/100$ (Table~\ref{tab:reliability}).

\begin{table}[t]
\centering
\small
\setlength{\tabcolsep}{4pt}
\begin{tabular}{lcccc}
\toprule
Metric & Apple & Gemma & Llama & Minis. \\
\midrule
Confidence AUROC $\uparrow$          & 0.47 & 0.47 & 0.52 & 0.50 \\
ECE $\downarrow$                     & \textbf{70.1} & 64.6 & 54.2 & 57.1 \\
Confident halluc.\ (\%) $\downarrow$ & \textbf{67.9} & 66.0 & 35.7 & 54.3 \\ 
FP confabulation (\%) $\downarrow$   & \textbf{69.1} & 26.4 & 12.7 & 7.3 \\ 
Over-refusal, benign (\%) $\downarrow$ & \textbf{18.4} & 0.0 & 0.0 & 0.0 \\
Grounded (\%) $\uparrow$             & \textbf{86.2} & 81.3 & 74.3 & 68.1 \\
\bottomrule
\end{tabular}
\caption{Reliability comparison against comparably sized small models on identical items with one grader. The on-device model is worst-in-class on calibration, confident confabulation, and over-refusal, yet most grounded on its intended task: the task-asymmetric pattern. Bold marks the worst (best for grounded). Peer comparisons are across each model's standard serving configuration (on-device guided decoding vs.\ hosted API for peers); serving-pipeline differences are a confound, so cross-model superlatives should be read as contextual. Our core claims (Sections~\ref{sec:surface}--\ref{sec:mitigation}) are within-model and do not depend on this contrast.}
\label{tab:reliability}
\end{table}

\subsection{Saturated, Miscalibrated Confidence}
\label{sec:calib}
\begin{figure}[t]
\centering
\includegraphics[width=0.95\columnwidth]{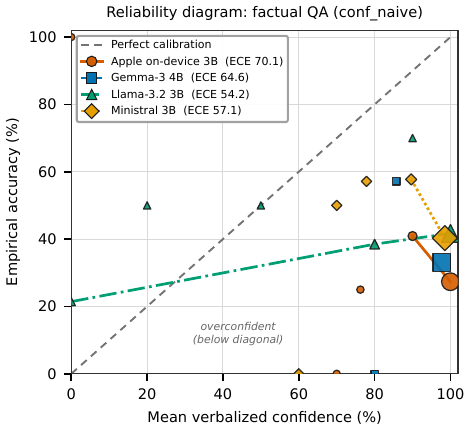}
\caption{Reliability diagram on factual QA. The on-device model's confidence sits near $100\%$ almost regardless of accuracy, far from the diagonal, yielding the worst ECE among comparable small models.}
\label{fig:reliability}
\end{figure}

On factual QA (TriviaQA and Global-MMLU), the model's verbalized confidence is almost constant (mean $98.8$ $[98.6, 98.9]$) and fails to separate correct from incorrect answers (AUROC $0.471$ $[0.456, 0.485]$, at or below chance). This saturation is not an artifact of decoding mode: mean verbalized confidence is $98.8$ under guided decoding and ${\sim}99$ under free-form decoding (Section~\ref{sec:costladder}), so the calibration and single-generation findings, though established under different decoding conditions, rest on the same saturated confidence signal. Its expected calibration error, $70.1$ $[68.4, 71.8]$, is the worst among the small models we test (Figure~\ref{fig:reliability}), exceeding Llama-3.2-3B by $+17.7$ points $[+15.2, +20.3]$, Ministral-3B by $+12.7$, and Gemma-3-4B by $+5.1$ (all $p<0.001$, paired). Because verbalized confidence is saturated near $99\%$, a large ECE is partly mechanical; we therefore treat \emph{discrimination} (AUROC $0.471$, at chance) as the substantive calibration signal, and read ECE as characterizing the verbalized-confidence channel specifically rather than an intrinsic property of the model's internal probabilities (which the on-device API does not expose). As a direct consequence, $67.9\%$ of its factual answers are delivered at high confidence yet wrong, a confident-hallucination rate $+32.2$ points $[+30.0, +34.3]$ above Llama ($n{=}2900$). These rankings are not a peer serving-pipeline artifact: re-evaluating all three peers under sampled decoding (temperature $0.7$) shifts their ECE, confident-hallucination, and confabulation rates by at most $3.5$ points and never overturns the ordering; the on-device model remains worst on all three under both greedy and sampled peer decoding (supplementary material, \S\,Peer Decoding Robustness).

\subsection{Confident Confabulation on False Premises}
\label{sec:fp}
When a question embeds a false premise (e.g., asking for a property of a nonexistent entity), the appropriate behavior is to abstain or correct the premise. Instead, on 110 such items the model abstains only $30.9\%$ $[22.7, 40.0]$ of the time and \emph{confidently confabulates} an answer $69.1\%$ $[60.0, 77.3]$ of the time. This confabulation rate exceeds every peer by a wide margin: $+42.7$ points over Gemma, $+56.4$ over Llama, and $+61.8$ $[+51.8, +71.8]$ over Ministral (all $p<0.001$, paired). The failure is specific to false premises: on 150 answerable control questions the model over-refuses only $0.7\%$ $[0.0, 2.0]$ of the time, confirming this is not blanket caution but a failure to recognize when a question should not be answered. Although individual rates carry moderate intervals (e.g., confabulation $69.1\%$ $[60.0, 77.3]$, $n{=}110$), the cross-model gaps ($+42$ to $+62$ points) far exceed those intervals, so the qualitative ordering is robust to sampling noise.

\subsection{Summarization: Grounded Yet Over-Refusing}
\label{sec:refusal}
\begin{figure}[t]
\centering
\includegraphics[width=0.95\columnwidth]{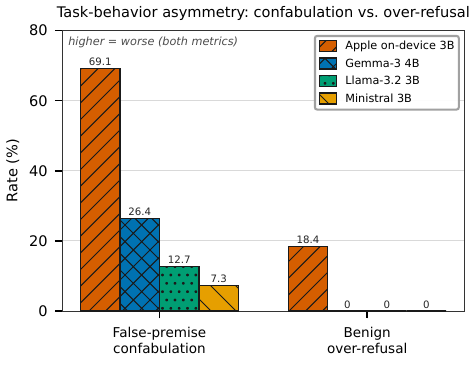}
\caption{Task-asymmetric miscalibration: the on-device model is high on \emph{both} opposite failures, confidently confabulating on false premises (under-abstention) and refusing benign inputs (over-abstention), while peers are high on neither.}
\label{fig:asymmetry}
\end{figure}

On its intended task, the model is competent: judged on 310 summarization items, it is the \emph{most grounded} small model we test ($86.2\%$ $[81.8, 90.3]$ grounded, $+12.0$ over Llama and $+18.2$ over Ministral, $p<0.001$ paired). Yet on the same benign inputs it \emph{refuses} $18.4\%$ ($57/310$) of the time, while every peer refuses $0.0\%$ ($\Delta +18.4$ $[+14.2, +22.9]$, $p<0.001$). This over-refusal is not a mis-set option: enabling Apple's own \texttt{permissiveContentTransformations} guardrail mode leaves it unchanged ($18.8\%\!\to\!18.4\%$; McNemar $p{=}1$; $1/304$ items changed). Together with Section~\ref{sec:fp}, this establishes the task-asymmetric miscalibration: too permissive where caution is warranted, too cautious where it is not (Figure~\ref{fig:asymmetry}).

\section{The Harm Mechanism: Surface Indistinguishability}
\label{sec:surface}
\begin{figure}[t]
\centering
\includegraphics[width=0.95\columnwidth]{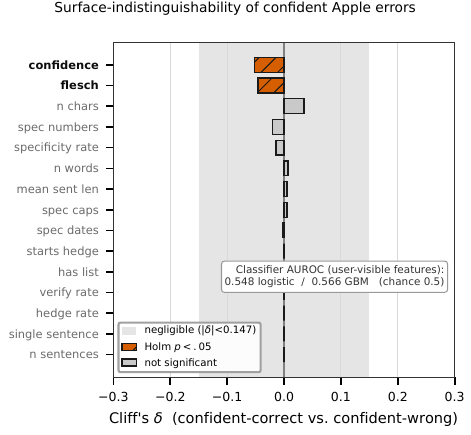}
\caption{Confident-correct vs.\ confident-wrong outputs are surface-indistinguishable: per-feature effect sizes fall inside the negligibility band ($|\delta|<0.147$), and a classifier using only user-visible features reaches AUROC $0.55$ (equivalent to chance by TOST).}
\label{fig:surface}
\end{figure}

For a user to guard against a confident error, confident-wrong outputs must \emph{look} different from confident-correct ones. They do not. On $N{=}3019$ high-confidence short-answer outputs (858 correct vs.\ 2161 wrong, including 108 false-premise confabulations), we compare 15 user-visible features (length, readability, hedging, specificity, stated confidence, etc.). We restrict attention to features a user or a resource-constrained developer can read off a single output: surface form (length, readability), lexical hedging, specificity markers, and the model's own stated confidence, rather than semantic signals that require extra model calls or ground truth; those are precisely what the consistency mitigation of Section~\ref{sec:mitigation} exploits. The largest effect size across all features is negligible ($\max|\text{Cliff's }\delta|=0.052$, well below the $0.147$ negligibility threshold); only stated confidence ($\delta{=}{-}0.052$) and Flesch readability ($\delta{=}{-}0.046$) survive Holm correction, both negligible. A classifier trained to predict correctness from user-visible features alone reaches AUROC $0.548$ $[0.525, 0.571]$ (logistic) and $0.566$ $[0.545, 0.588]$ (gradient boosting), barely above chance against an oracle of $1.000$ (Figure~\ref{fig:surface}). A TOST equivalence test confirms this is statistically indistinguishable from chance: all 15 per-feature contrasts are $\delta$-equivalent and both classifiers' upper $90\%$ bounds fall below $0.60$. We set the equivalence margin at AUROC $0.60$ as the minimum discrimination a downstream filter would need to be even marginally useful; an upper bound below $0.60$ thus certifies that no user-visible feature set reaches practical utility. Notably, the model's confident errors are \emph{less} surface-detectable than any peer's (Gemma $0.542$, Llama $0.591$, Ministral $0.641$). There is, in short, no user-visible tell.

\section{Can Cheap Signals Flag the Confident Failures?}
\label{sec:costladder}
Given that users cannot detect these errors (Section~\ref{sec:surface}), we ask whether an inexpensive, black-box signal could. We evaluate four single-generation, $O(1)$ signals: verbalized confidence, a hedge-to-verify ratio (HVR) over reasoning traces~\citep{selfdoubt}, their $z$-score fusion, and raw response length. We compare these against one $O(N)$ signal, $k$-sample self-consistency (SelfCheck, $k{=}5$), on two free-form tasks (factual TriviaQA and reasoning GSM8K) with frozen, disjoint calibration/evaluation splits, scoring every signal against the same free-form correctness label (Table~\ref{tab:o1o2}).

\paragraph{No single-generation signal is robust and task-general.}
Verbalized confidence is saturated (mean ${\sim}99\%$; cf.\ Section~\ref{sec:calib}) and non-discriminative (AUROC $0.529$ / $0.519$). Trace-based HVR is at chance on factual QA ($0.543$ $[0.463, 0.621]$) and only weakly above chance on reasoning ($0.590$ $[0.523, 0.657]$); three converging length-control tests show it adds \emph{nothing} beyond response length (partial-coefficient $p{=}0.69$/$0.98$; likelihood-ratio test $p{=}0.69$/$0.98$; length-residualized HVR at chance, $0.449$ $[0.379, 0.519]$ and $0.502$ $[0.423, 0.582]$). Response length is the only cheap signal with any power, and it is \emph{task-dependent}: at chance on factual QA ($0.513$) but moderately informative on reasoning ($0.680$), consistent with a difficulty proxy that exists only when the model deliberates. This reverses the pattern for server-side reasoning models, where HVR \emph{exceeds} length~\citep{selfdoubt}.

\paragraph{Only $k$-sample consistency discriminates robustly.}
SelfCheck beats the \emph{strongest} $O(1)$ baseline on both tasks: versus fused confidence on factual QA, $\Delta{+}0.194$ (DeLong $p{=}1.3\mathrm{e}{-}4$), and versus response length on reasoning, $\Delta{+}0.108$ (DeLong $p{=}2.3\mathrm{e}{-}3$). The reasoning-task margin over free response length is modest but significant, so we treat consistency sampling as a \emph{tunable} layer, not an always-on requirement (Section~\ref{sec:mitigation}).

\begin{table}[t]
\centering
\small
\begin{tabular}{lcc}
\toprule
Signal & TriviaQA & GSM8K \\
\midrule
\multicolumn{3}{l}{\emph{$O(1)$ single-generation}}\\
Verbalized confidence & 0.529 & 0.519 \\
HVR                    & 0.543 & 0.590 \\
Fused (HVR+conf)       & 0.555 & 0.598 \\
Response length        & 0.513 & \textbf{0.680} \\
\midrule
\multicolumn{3}{l}{\emph{$O(N)$ sampling}}\\
SelfCheck ($k{=}5$)    & \textbf{0.749} & \textbf{0.789} \\
\bottomrule
\end{tabular}
\caption{Correct-vs-incorrect discrimination (AUROC, free-form correctness label). No $O(1)$ signal is robust and task-general; only $O(N)$ consistency beats the strongest cheap baseline on both (DeLong $p<0.003$). Full 95\% CIs in the supplementary material.}
\label{tab:o1o2}
\end{table}

\section{Mitigation: Black-Box Recovery Without Model Access}
\label{sec:mitigation}
Can a third-party developer, with only black-box query access, recover reliability without retraining or privileged access? Our mitigation applies self-consistency~\citep{wang2023selfcons} in the black-box, zero-resource form of SelfCheckGPT~\citep{manakul2023selfcheck}: for each query we draw $k$ stochastic samples and abstain (or escalate) when their final answers disagree, converting undetectable confident errors into explicit abstentions. The method is not itself novel; our contribution is to show that it \emph{recovers} reliability on the deployed on-device model using only black-box query access (no logits, no fine-tuning, and no model internals), while running on device.

\begin{figure}[t]
\centering
\includegraphics[width=0.95\columnwidth]{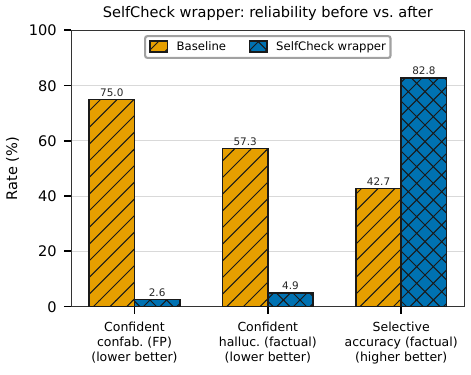}
\caption{Black-box SelfCheck ($k{=}5$) recovery: confident confabulation and hallucination collapse while selective accuracy nearly doubles, with no model access.}
\label{fig:mitigation}
\end{figure}

At $k{=}5$ (threshold tuned on held-out dev items), on the false-premise set it reduces confident confabulation from $75.0\%$ to $2.6\%$ ($\Delta{-}72.4$ $[-81.6, -61.8]$, $n{=}76$), raising abstention to $97.4\%$; on factual QA it cuts confident hallucination from $57.3\%$ to $4.9\%$ ($n{=}103$) and raises selective accuracy from $42.7\%$ to $82.8\%$ (ECE $57.2\!\to\!17.2$) at $28.2\%$ coverage. As a ranking signal, SelfCheck agreement separates correct from wrong at AUROC $0.77$ (TriviaQA) and $0.87$ (GSM8K) on its native plurality label, far above any single-generation signal (Figure~\ref{fig:mitigation}).

\begin{figure}[t]
\centering
\includegraphics[width=0.95\columnwidth]{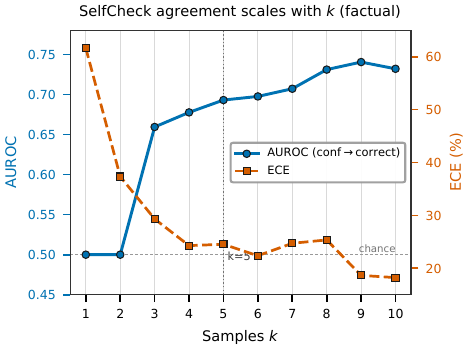}
\caption{Consistency-gated abstention: discrimination and calibration improve with the number of samples $k$ and are largely captured by $k{=}5$ (marked), recovering reliability without model access.}
\label{fig:kscale}
\end{figure}

\paragraph{Cost is tunable, not fixed.}
The wrapper costs $k\times$ generation, a genuine on-device expense in battery and latency, but it is a knob: a modest $k$ (we use $k{=}5$) captures most of the available signal (Figure~\ref{fig:kscale}), and the accept/defer threshold $\tau$ traces a full accuracy--coverage frontier (supplementary material, \S\,Risk--Coverage Frontier), so the wrapper is not tied to the single operating point reported here; escalation can be gated to high-stakes or low-agreement queries, and because the samples are independent, on-device batching keeps wall-clock latency below the nominal $k\times$ factor. More fundamentally, the wrapper shows the documented failures are \emph{recoverable without any privileged access to the model}.

\section{Discussion}
\label{sec:discussion}

\paragraph{Task-asymmetric miscalibration is an oversight failure, not a tuning bug.}
The two guardrail failures we document point in opposite directions, under-abstention on false premises (Section~\ref{sec:fp}) and over-refusal on benign inputs (Section~\ref{sec:refusal}), yet both survive the vendor's own permissive guardrail mode (Section~\ref{sec:refusal}). This asymmetry is hard to explain as a single mis-set threshold: a more permissive setting would worsen confabulation, while a more conservative one would worsen over-refusal. The model does not have a calibrated notion of \emph{when} to withhold an answer, and its saturated confidence signal (Section~\ref{sec:calib}) gives the surrounding application nothing to gate on. For a system deployed without server-side moderation, this is precisely the regime in which scalable oversight is supposed to help, and precisely where the model's self-reported signal fails.

\paragraph{Why undetectability is the load-bearing finding.}
A confidently wrong answer is tolerable if the user can recognize it as suspect and verify it. Our surface-indistinguishability result (Section~\ref{sec:surface}) removes that escape hatch: across fifteen user-visible features, confident-correct and confident-wrong outputs are statistically equivalent, and the model's errors are in fact \emph{less} surface-detectable than any peer's. We are careful about the scope of this claim: it establishes a \emph{necessary condition} for user self-protection to fail (there is no usable cue in the output), not a measured rate of downstream harm, which would require a user study. But the necessary condition is the part a developer or user cannot fix at inference time, and it is what turns a calibration problem into a safety problem.

\paragraph{The accountability gap: benchmarked is not shippable.}
The configuration a vendor benchmarks and the configuration a developer can deploy are not the same object. Capability and safety claims made for a private, task-specialized adapter do not transfer to the general base model third parties actually ship, and the base model cannot be independently audited against those claims because the adapter is not exposed. This is an accountability gap: the publicly defensible number describes a system the public cannot use or verify. We do not read this as vendor misconduct (the private configuration may well perform as reported) but as a structural problem for evaluation and governance. Independent audits, and the claims that inform policy, should target the deployable configuration explicitly.

\paragraph{Implications for deployment and evaluation.}
Our cost-ladder (Section~\ref{sec:costladder}) shows that the cheap, single-generation uncertainty signals that work on server-side reasoning models do not transfer to this deployed on-device setting: self-reported confidence is saturated, hedging adds nothing beyond response length, and length itself is informative only when the model genuinely reasons. For models exhibiting this confidence saturation, practitioners should not rely on a single forward pass to know when the model is wrong. The good news is that the failures are \emph{recoverable} by a black-box consistency wrapper that needs no model access (Section~\ref{sec:mitigation}), at a cost that is tunable rather than fixed. This suggests a concrete design pattern for on-device applications in high-stakes settings, consistency-gated abstention, and a concrete recommendation for evaluators: audit the deployable model on reliability axes (calibration, confabulation, over-refusal, and detectability), not capability alone.

\section{Limitations}
\label{sec:limitations}
Our study has several limitations. First, it audits a \emph{single} deployed model from one vendor; while the audit methodology, the surface-indistinguishability protocol, and the black-box wrapper are model-agnostic, the specific rates we report do not necessarily generalize to other on-device models. The methodology, however, is not single-model: the calibration, surface-indistinguishability, and cost-ladder analyses are applied to all four models here (e.g., the surface protocol yields classifier AUROCs of $0.54$--$0.64$ on the peers, Section~\ref{sec:surface}), so transportability of the \emph{methods} is demonstrated, not merely conjectured; the released code and frozen items let others run the same audit on any model. Second, our claims are scoped to the developer-accessible configuration as exposed by the public FoundationModels API at the time of study; the underlying build and operating-system version are not recorded in our generation logs, and on-device model behavior may change across updates; we will pin the exact build for the camera-ready. Third, summarization groundedness relies on an LLM judge (audited at $88\%$ agreement by a second judge) rather than a human audit; the retained human-label sample is left for future work. Fourth, the single-generation uncertainty probe covers two task types (open-domain factual QA and multi-step reasoning); other task families, and multilingual behavior beyond our sampled items, remain untested. Fifth, our harm argument establishes a necessary condition, the absence of any user-visible cue, rather than a measured rate of downstream harm, which would require a human-subjects study. Finally, peers are served through a hosted API rather than on device; although items, prompts, and grading are matched, serving-pipeline differences cannot be fully excluded, and the mitigation is demonstrated at a single tuned operating point rather than swept across all deployment budgets.

\section{Conclusion}
\label{sec:conclusion}
We audited the developer-accessible configuration of a widely deployed on-device language model and found a task-asymmetric miscalibration (confident confabulation where caution is warranted, over-refusal where it is not) built on a confidence signal that is saturated and non-discriminative. Crucially, its confident errors are surface-indistinguishable from its confident-correct outputs, leaving users no cue to detect them, and no cheap single-generation signal recovers that cue; only $O(N)$ consistency does, via a black-box wrapper that needs no model access. Two lessons extend beyond this specific model: reliability audits should target the \emph{deployable} configuration rather than only the benchmarked one; and, at least for the model we study (and, we conjecture, for similar deployed models), oversight cannot rely on a single forward pass. We release our code and frozen items to support both.

\section*{Ethical Statement}
This work is an independent reliability audit of a widely deployed, developer-accessible language model, conducted entirely through the public API using public benchmark data and no human subjects. Our intent is to improve oversight of deployed on-device models, by documenting failure modes that reach users without server-side moderation and by providing a black-box mitigation, not to disparage any vendor; we report findings as measurements with confidence intervals and explicitly do not contest the vendor's results for its private configurations. We recognize the dual-use character of red-teaming: surfacing confident-confabulation and over-refusal behaviors could inform misuse. We judge the disclosure benefit to substantially outweigh this risk, because the behaviors are readily observed by any developer, affect end users directly, and are accompanied here by a concrete, access-free mitigation. All code and frozen evaluation items are released to support independent verification and to encourage auditing of deployable, rather than only benchmarked, model configurations.

\bibliography{refs}

\end{document}